\documentclass[prd,aps,showpacs,nofootinbib,onecolumn,superscriptaddress,
amssymb]{revtex4}

\usepackage{graphicx}
\usepackage[english]{babel}
\usepackage{amsmath}
\usepackage{amssymb}
\usepackage{amsfonts}
\usepackage{latexsym}
\usepackage{color,amsxtra}
\usepackage{epsf}
\usepackage{enumerate}
\usepackage{hhline}
\usepackage{array}
\usepackage{tabularx}
\usepackage{subfigure}
\usepackage{fancyhdr}
\usepackage{mathrsfs}
\newcommand{\be}{\begin{equation}}
\newcommand{\ee}{\end{equation}}
\newcommand{\bea}{\begin{eqnarray}}
\newcommand{\eea}{\end{eqnarray}}
\newcommand{\beaa}{\begin{eqnarray*}}
\newcommand{\eeaa}{\end{eqnarray*}}

\def\be{\begin{equation}}
\def\ee{\end{equation}}
\def\bea{\begin{eqnarray}}
\def\eea{\end{eqnarray}}

\begin{document}

\title{Cosmological Fluids with Logarithmic Equation of State}

\author{Sergei D. Odintsov}
\email{odintsov@ieec.uab.es}
\affiliation{Institut de Ciencies de lEspai (IEEC-CSIC),
Campus UAB, Carrer de Can Magrans, s/n
08193 Cerdanyola del Valles, Barcelona, Spain}
\affiliation{Instituci\'{o} Catalana de Recerca i Estudis Avan\c{c}ats
(ICREA), Passeig Llu\'{i}s Companys, 23 08010 Barcelona, Spain}

\author{V.K. Oikonomou}
\email{v.k.oikonomou1979@gmail.com} \affiliation{Department of
Physics, Aristotle University of Thessaloniki, Thessaloniki 54124,
Greece}\affiliation{Laboratory for Theoretical Cosmology, Tomsk
State University of Control Systems and Radioelectronics, 634050
Tomsk, Russia (TUSUR)} \affiliation{Tomsk State Pedagogical
University, 634061 Tomsk, Russia}

\author{A.V. Timoshkin}
\email{alex.timosh@rambler.ru} \affiliation{Tomsk State
Pedagogical University, Kievskaja Street, 60, 634061 Tomsk,
Russia} \affiliation{ National Research Tomsk State University,
Lenin Avenue, 36, 634050 Tomsk, Russia}

\author{Emmanuel N. Saridakis}
\email{Emmanuel\_Saridakis@baylor.edu}
\affiliation{Department of Physics, National Technical University of Athens, Zografou
Campus GR 157 73, Athens, Greece}
\affiliation{CASPER, Physics Department, Baylor University, Waco, TX 76798-7310, USA}
\affiliation{ Eurasian  International Center for Theoretical
Physics, Eurasian National University, Astana 010008, Kazakhstan}

\author{R. Myrzakulov}\email{rmyrzakulov@gmail.com}
\affiliation{ Eurasian  International Center for Theoretical
Physics, Eurasian National University, Astana 010008, Kazakhstan}


\begin{abstract}
We investigate the cosmological applications of fluids having an equation of state which
is the analog to the one related to the isotropic deformation of crystalline solids, that
is containing logarithmic terms of the energy density, allowing additionally for a bulk
viscosity. We consider two classes of scenarios and we show that they are both capable of
triggering the  transition from   deceleration to acceleration at late times.
Furthermore, we confront the scenarios with data from Supernovae type
Ia (SN Ia) and Hubble function observations, showing that the agreement is excellent.
Moreover, we perform a
dynamical system analysis and we show that there exist asymptotic accelerating
attractors, arisen from the logarithmic terms as well as from the viscosity, which in most
cases correspond to a phantom late-time evolution. Finally, for some parameter regions
we obtain a nearly de Sitter late-time attractor, which is a significant
capability of the scenario since the dark energy, although dynamical, stabilizes at the
cosmological constant value.
\end{abstract}


\pacs{04.50.Kd, 95.36.+x, 98.80.-k, 98.80.Cq,11.25.-w}

\maketitle

\section{Introduction}

One of the main aims of modern cosmology is to explain the current
accelerated expansion of the Universe
\cite{Riess:1998cb,Perlmutter:1998np}. According to
cosmological observations at present, approximately 70\% of the energy
density of the Universe
is attributed to a component dubbed dark energy
\cite{Copeland:2006wr,Cai:2009zp}. The remaining
26\% is attributed to cold dark matter (CDM) and only 4\%
corresponds to ordinary baryonic matter. Dark matter is necessary in order to provide an
explanation to the peculiar velocities in galaxies as well as to the cluster collisions.
Although dark matter may be some sort of non-interacting particle
\cite{Oikonomou:2006mh}, nowadays there exist various alternative proposals
mimicking it
\cite{Dutta:2017fjw,Sebastiani:2016ras,Nashed:2018aai}.

In   $\Lambda$CDM paradigm dark matter is modeled as a pressureless
fluid, while  dark energy is assumed to not interact with ordinary matter and
it can be interpreted as  the vacuum energy, namely a simple cosmological constant. On
the other hand, in general dark energy models the accelerating expansion can be
described in terms of an exotic perfect fluid with negative pressure, which satisfies a
barotropic equation of state
\cite{Peebles:2002gy,Sahni:1999gb,Li:2011sd}. Hence, many
studies in the literature use perfect fluids to describe various
evolutionary aspects of the Universe
\cite{Barrow:1994nx,Tsagas:1998jm,HipolitoRicaldi:2009je,Gorini:2005nw,Kremer:2003vs,
Carturan:2002si,Brevik:2017juz,Buchert:2001sa,Hwang:2001fb,Elizalde:2009gx,
Cruz:2011zza,Oikonomou:2017mlk,
Brevik:2017msy,Capozziello:2006dj,Elizalde:2017dmu,
Brevik:2016kuy,
Balakin:2012ee,Brevik:2018azs}, while
the most general
models of dark fluids can be incorporated using an inhomogeneous
equation of state
\cite{Nojiri:2005sr,Nojiri:2006zh,Elizalde:2005ju}.
Finally, an alternative description of the dark energy sector can arise effectively
through gravitational modifications
\cite{reviews1,reviews2,reviews3,reviews4,reviews4b,reviews5,reviews6,reviews7}.

As it is known, the standard cold dark matter scenario provides
very efficient results at large (cosmological) scales, but it might be
problematic at galactic scales. These problems may be connected to the
assumption that dark matter is pressureless. Nevertheless, the description of
the late-time Universe at small scales can be achieved through the framework of the
logarithmic-corrected equation of state for the
matter sector, within the Debye approximation
\cite{Intermetallics1,Intermetallics2}. In this formulation  the fluid pressure is modeled
by an empirical formula for the pressure of the deformed crystalline solids under
the isotropic stress \cite{Ivanovskii}. In order for the Universe
to change under the action of cosmic expansion, it is necessary
that the fluid pressure, described in terms of an equation of
state, to be negative \cite{reviews1,reviews2,reviews3}. The
negative pressure in the logarithmic-corrected equation of state scenarios
 becomes dominant when the volume of the Universe exceeds a
certain value. This scenario corresponds to the approaches of
logotropic dark energy model (LDE) \cite{Chavanis:2014hba,Chavanis:2015eka}.

In the present work we are interested in studying the dynamical evolution of a
late-time Universe by assuming a modified log-corrected power-law
equation of state (EoS) for the dark energy fluid, allowing additionally the fluid to be
viscous.  In particular, we  examine in
detail various forms of the dark energy   EoS and we
investigate how the Universe evolution   is affected by the
corresponding EoS. Moreover, we shall assume that a non-trivial
interaction between the dark energy and dark matter fluids may exist,
and by conveniently choosing the EoS we shall investigate the
dynamical behavior of the cosmological system in terms of an
autonomous dynamical system. The motivation for using non-trivial
interaction between the dark sectors arises mainly from the fact
that dark energy dominates over dark matter at late times, and thus  it is possible
that the dark matter sector loses
its energy feeding the dark energy sector and providing an alleviation to the coincidence
problem
\cite{Gondolo:2002fh,Farrar:2003uw,Cai:2004dk,Wang:2006qw,
Bertolami:2007zm,Chen:2008ft,
He:2008tn,Valiviita:2008iv,Jackson:2009mz,Jamil:2009eb,He:2010im,
Boehmer:2008av,Bamba:2012cp,Bolotin:2013jpa,
Costa:2013sva,Nunes:2016dlj,Odintsov:2018awm,Odintsov:2018uaw}. The resulting
picture of the coupled dark
energy - dark matter system is quite interesting as we demonstrate,
since for a class of parameter values the dynamical system has a
stable asymptotic attractor, which corresponds to  an accelerating phantom
fixed point. Notably, this phantom
attractor can become a nearly de Sitter attractor if the model
parameters are chosen appropriately.

This paper is organized as follows: In section \ref{ModifiedEquations} we present in
brief the motivation and the structure of the proposed EoS. In section
\ref{ViscousLogarithmic}  we discuss the
cosmological applications of two scenarios, with emphasis given in
the late-time era. In section \ref{Observational} we briefly discuss the
observational constraints and implications of the logarithmic
corrected EoS dark energy fluid. In section \ref{DynamicalSystem} we apply the
dynamical system method in order to investigate the
coupled dark energy-dark matter system, with the dark energy
sector being described by a viscous logarithmic EoS of a specific
form.   Finally, the
conclusions follow at the end of the paper.

\section{Viscous logarithmic-corrected power-law fluid}
\label{ModifiedEquations}

Our purpose is to study dark energy in terms of a
logarithmic-corrected power-law fluid. The EoS of such a fluid has
the form \cite{Capozziello:2017buj}
\begin{equation}
\label{Eq1}
 p=A\left(\frac{\rho}{\rho^*}\right)^{-l}\ln \left(\frac{\rho}{\rho^*}\right),
\end{equation}
where $\rho^*$ is a reference density, which is identified with
the Planck density in \cite{Chavanis:2014hba}, namely $\rho_{Pl}=c^5/(\hbar
G)\approx 5.16\times 10^{99}gr/m^3$. In the new notation $A>0$
represents the logotropic temperature, while
$l=-\frac{1}{6}-\gamma_G$, with $\gamma_G$ the dimensional Gruneisen parameter. For $l=0$
we obtain the equation of
state for the logotropic cosmological model
\cite{Chavanis:2014hba}. It is interesting that the EoS of Eq.
(\ref{Eq1}) may have deep relation with equations of state found
in condensed matter fluids, introduced in Refs.
\cite{Intermetallics1,Intermetallics2}.

Let us rewrite   (\ref{Eq1}) in the notation of a logotropic dark energy model LDE
model. For this purpose we express the volume in terms of mass
density, using the relation
\cite{Chavanis:2014hba,Chavanis:2015eka}
\begin{equation}
\label{Eq2}
 p(V)=-\beta\left(\frac{V}{V_0}\right)^{-\frac{1}{6}-\gamma_G}\ln
\left(\frac{V}{V_0}\right),
\end{equation}
where $V_0$ is a volume, which presents a barrier among the
different signs of the pressure $p$, and $\beta$  is a bulk modulus at
$V_0$. The
bulk modulus shows how much the volume changes under the action of
external forces. The parameter  $\gamma_G$  in the homogeneous and
isotropic Universe is a free parameter of the theory. When $V<V_0$
the pressure is positive for the positive bulk modulus, and it is
negative in the case of an inequality of the opposite sign. If the
pressure of the dark fluid satisfies relation (\ref{Eq2}),
then in order to ensure the cosmic acceleration
the volume must overcome the barrier $V\approx V_0$. There are
three different regimes of the behavior of the pressure
(\ref{Eq2}) \cite{Capozziello:2017buj}:

\begin{enumerate}
    \item The era before passing the $V_0$  barrier, when  $V<V_0$. Then the pressure is
positive and the Universe is decelerating. This case corresponds
to the case of the pressureless matter in the $\Lambda$CDM model.
    \item The era of equivalence between volumes, when  $V=V_0$. That is the transition
time
from the deceleration    to the acceleration era.
    \item The era after passing the $V_0$  barrier, when  $V>V_0$. Then the pressure is
negative and the fluid starts to trigger the Universe acceleration. Thus, in
the logarithmic-corrected power-law model, the dynamical evolution
of the Universe is described by a single fluid, which accelerates
the Universe, when its volume passes the barrier $V=V_0$. This
allows us to apply this model to the description of the late
Universe.
\end{enumerate}

We consider a homogeneous and isotropic on large scales flat geometry,  and we desire
to study the dynamical evolution of the Universe using a
fluid described by the logarithmic EoS (\ref{Eq1}). For generality, we additionally
allow for  viscosity of the fluid. In order to achieve this we
modify equation (\ref{Eq1}) adding the term which describes
viscosity, namely
\begin{equation}
\label{Eq3}
\zeta(H,t)=\xi_1(t)(3H)^n,
\end{equation}
 where $\zeta(H,t)$ is the bulk viscosity which depends on the Hubble parameter  $H$
and on the time $t$. From thermodynamic considerations it follows
that $\zeta(H,t)>0$.

In summary, in view of the above considerations, the EoS
for the logarithmic-corrected power-law fluid has the form
\begin{equation}
\label{Eq4}
 p=A\left(\frac{\rho}{\rho^*}\right)^{-l}\ln \left(\frac{\rho}{\rho^*}\right)-3H
\zeta(H,t).
\end{equation}
 In the next
section we shall investigate the   evolution of the
Universe for various forms of the dark energy EoS.

\section{Cosmology with viscous logarithmic-corrected power-law fluid}
\label{ViscousLogarithmic}

In this section we will study the late-time behavior of the
Universe by using an inhomogeneous viscous fluid description of
dark energy. We consider a spatially flat Friedmann-Robertson-Walker (FRW)
space-time with line element
\begin{eqnarray}
ds^2=-dt^2+a(t)^2\delta_{ij}dx^i dx^j\, ,
\end{eqnarray}
where $a(t)$ is the scale factor.

  The
logarithmic-corrected power-law fluid model cannot describe the
early Universe, because the temperature during this era is much
greater than the Debye temperature of solids. The fluid in the
inflationary epoch becomes pressureless as in the case of LDE
model \cite{Chavanis:2016pcp}, while at   late times the pressure
tends to a constant negative value and thus it provides the necessary requirement
for triggering the acceleration. Additionally, the incorporation of
viscosity improves the  singularity structure of the cosmological
system at hand, and the behavior of the Universe in the vicinity
of a Big Rip  \cite{Caldwell:1999ew,Caldwell:2003vq,Faraoni:2001tq,Nojiri:2003vn}
or of types II, III and IV singularities
\cite{Barrow:2004xh,Nojiri:2004ip}, the classification of which
was first given in \cite{Nojiri:2005sx}.

Let us first consider the non-interacting scenario, in which case $\rho$ fulfills the
standard conservation equation
\begin{equation}\label{Eq5}
\dot{\rho}+3H(\rho+p)=0\, ,
\end{equation}
where $H=\dot{a}/a$  is the Hubble function, and $p$ is given by (\ref{Eq4}). In the
following two subsections we study two scenarios with different fluid viscosity form.

\subsection{Viscosity with constant function $\xi_1(t)$}

We start by considering the bulk viscosity in the expression (\ref{Eq3})
to have the following simple form:
\begin{equation}\label{Eq9}
\zeta (H,t)=3\alpha H\, ,
\end{equation}
where $\alpha$  is a positive constant, since such a linear-in-$H$ form is widely used in
viscous cosmology \cite{Brevik:2017msy}. Hence, inserting (\ref{Eq4}) and (\ref{Eq9})
into (\ref{Eq5}), and using the Friedmann equation we obtain
\begin{equation}\label{Eq10}
\dot{\rho}+3H\Big{[}A\left(\frac{\rho}{\rho_*} \right)^{-l}\ln
\frac{\rho}{\rho_*}+\theta \rho\Big{]}=0\, ,
\end{equation}
where  $\theta=1+3\alpha \kappa^2$,
with $\kappa^2=8\pi G$ and $G$ denoting the Newton's gravitational constant.

In the low-energy regime
($\rho\ll\rho_*$), and for  $l=-1$, equation (\ref{Eq10}) simplifies as
\begin{equation}\label{Eq11}
\dot{\rho}+\frac{3H}{\rho_*}\Big{[}A(\rho-\rho_*)+\theta\rho\rho_*\Big{]}=0\,,
\end{equation}
which using the second Friedmann equation
becomes
\begin{equation}\label{Eq12}
\dot{H}+d\,H^2-b=0\, ,
\end{equation}
where $d=\frac{3}{2}(A+\theta\rho_*)$  and
$b=\frac{1}{2}A\kappa^2\rho_*$ (note that $b>0$ since $A>0$). The solution of   equation
(\ref{Eq12}) is found to be
\begin{equation}\label{Eq13}
H(t)=\sqrt{\frac{b}{d}}\frac{e^{\sqrt{bd}t}+C_1e^{\sqrt{bd}t}}{e^{\sqrt{bd}t}-C_1e^{\sqrt{
bd}t}}\,,
\end{equation}
where $C_1$  is an arbitrary constant. As we observe, the Hubble function $H(t)$
diverges for $t\to 0$ and a Big Bang type singularity occurs.
Additionally, taking $C_1=1$ for simplicity, the scale factor is given by the expression
\begin{equation}\label{Eq14}
a(t)=a_0\sinh(\sqrt{bd}t)^{1/d}\, ,
\end{equation}
with $a_0$ an integration constant,
while its second derivative reads as
\begin{equation}\label{Eq15}
\ddot{a}=\frac{b}{d}\left[\frac{\cosh^2
\left(\sqrt{bd}t\right)-d}{\sinh^2\left(\sqrt{bd}t\right)}\right]a(t)\,.
\end{equation}
Hence, we obtain $\ddot{a}=0$  at $t_0=\frac{1}{\sqrt{bd}}\ln
\left(\sqrt{d}+\sqrt{d-1}\right)$. If $d<1$, or equivalently if
$A<\frac{2}{3}-\theta \rho_*$, then the second derivative of the
scale factor is positive and the Universe transits to an
accelerated expansion.
On the other hand, in the case $d>1$, then for $0<t<t_0$ the
second derivative is negative, i.e.  the expansion is
decelerating, while for $t>t_0$  the Universes enters in an
accelerating era. Therefore, we are able to obtain the
 transition from a decelerating to an accelerating epoch. Finally, note that from
(\ref{Eq14}) we
find that
 \begin{equation}\label{Eq16}
\dot{H}(t)=-\frac{b}{\sinh^2 \left( \sqrt{bd}t\right)}\, ,
\end{equation}
and thus we deduce that since $\dot{H}(t)<0$  we obtain a Universe that
does not super-accelerate. Lastly, we mention here that in the
case of zero viscosity (i.e. for $\alpha=0$) we obtain $\theta=1$
and thus the above analysis is significantly simplified.

In summary, the model at hand can describe the Universe evolution,
with transition from deceleration to   acceleration epoch. We
mention that we have not considered an explicit cosmological
constant, and thus the above behavior arises only due to the model
dynamics.


\subsection{Viscosity with a linear
time-dependent function $\xi_1(t)$ }

Let us now assume that the function $\xi_1(t)$ in (\ref{Eq3}) has the form
\begin{equation}\label{Eq17}
\xi_1(t)=d_1t+b_1,
\end{equation}
with $d_1$  and $b_1$ being arbitrary parameters, that is we consider the bulk viscosity
to have the following simple form:
\begin{equation}\label{Eq17b}
\zeta (H,t)=3 H(d_1t+b_1)\, .
\end{equation}
Thus, inserting (\ref{Eq4}) and
(\ref{Eq17b})
into (\ref{Eq5}), and using the Friedmann equation we acquire
\begin{equation}\label{Eq18}
\dot{\rho}+3H\Big{[}A\left(\frac{\rho}{\rho_*}\right)^{-l}-A\left(\frac{\rho}{\rho_*}
\right)^{-l-1}
+(\tilde{c}t+\tilde{b})\rho\Big{]}=0\,,
\end{equation}
where $\tilde{c}=3d_1\kappa^2$ and $\tilde{b}=1-3b_1\kappa^2$. In the case
$l=-1$ , using the second Friedmann equation  we can rewrite
(\ref{Eq18}) as
\begin{equation}\label{Eq19}
2\dot{H}+3\left(\tilde{c}t+\tilde{b}+\frac{A}{\rho_*}
\right)H^2=0\, ,
\end{equation}
which has the solution
\begin{equation}\label{Eq20}
H(t)=\frac{4}{3\left(\tilde{c}t+\tilde{b}+\frac{A}{\rho_*}
\right)^2+C_2}\, ,
\end{equation}
with $C_2$   an integration constant. Without loss of generality
we focus on the case $C_2=0$. Firstly, we can see that the Hubble rate $H(t)$ diverges
at the finite time
$t_0=-\frac{1}{\tilde{c}}\left(\tilde{b}+\frac{A}{\rho_*} \right)$, and thus a Big Rip
type singularity occurs. Concerning the scale
factor, solution (\ref{Eq20}) leads to
\begin{equation}\label{Eq21}
a(t)=a_0\exp\Big{[}-\frac{4}{3\tilde{c}}\left(\tilde{c}t+\tilde{b}+\frac{A}{\rho_*}
\right)^{-1} \Big{]}\, ,
\end{equation}
with $a_0$ an integration constant,
while its second derivative is
\begin{equation}\label{Eq22}
\ddot{a}(t)=\Big{[}1-\frac{3}{2}\tilde{c}\left(\tilde{c}t+\tilde{b}+\frac{A}{\rho_*}
\right) \Big{]}H^2(t)a(t)\, .
\end{equation}
As we observe, $\ddot{a}=0$  at
$t_1=\frac{1}{\tilde{c}}\Big{[}\frac{2}{3\tilde{c}}-\tilde{b}-\frac{A}{\rho_*}\Big{]}$.
Thus, in the case $\tilde{c}>0$, for values $t<t_1$, it turns out
that $\ddot{a}<0$, and therefore the Universe experiences a decelerated
expansion, while for $t>t_1$ we have $\ddot{a}>0$  and the
Universe transits to a late-time accelerated era.

In summary, the model at hand can describe the Universe's evolution, with the transition
from decelerating   to   accelerating epoch. We mention that this behavior is obtained
although we have not considered an explicit cosmological constant.

\section{Observational Constraints}
\label{Observational}

As we saw, the scenario at hand can give rise to a Universe that
experiences the transition from a decelerating era to an
accelerating epoch. In order to perform a confrontation with
observations we will use the Supernovae type Ia (SN Ia) data, as well as direct
measurements of the Hubble parameter
with the corresponding covariance matrix.

\subsection{Supernovae Type Ia:}

  In these observational data sets,  the apparent luminosity  $l(z)$
(or equivalently the apparent magnitude $m(z)$), is measured as a
function of the redshift, and is related to the luminosity
distance as
\begin{equation}
2.5 \log\left[\frac{L}{l(z)}\right] = \mu \equiv m(z) - M = 5
\log\left[\frac{d_L(z)_{\text{obs}}}{Mpc}\right]  + 25,
\end{equation}
where $M$ is the absolute magnitude  and $L$ is the luminosity.
Moreover, for any  model under consideration one can calculate the
theoretically predicted dimensionless luminosity distance
$d_{L}(z)_\text{th}$ using the theoretically predicted evolution
of the Hubble function through
\begin{equation}
d_{L}\left(z\right)_\text{th}\equiv\left(1+z\right)
\int^{z}_{0}\frac{dz'}{H\left(z'\right)}~.
\end{equation}
\begin{figure}[ht]
\includegraphics[width=9.4cm,height=7cm]{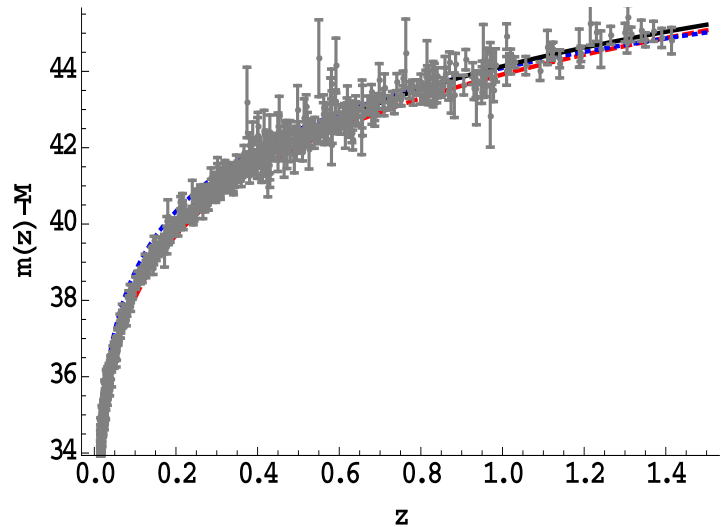}
\caption{ {\it{The theoretically
predicted apparent minus absolute magnitude as a function of the
redshift, for the logarithmic-corrected    fluid  (\ref{Eq4}) with   viscosity
 (\ref{Eq9}) (i.e.  constant function $\xi_1(t)$), for $a_0=1$ and $b=0.25\times
10^{-120}$,
 in units
where
$\kappa^2 = c = \hbar = 1$ (i.e where $H(z=0)\equiv H_0\approx 6
\times 10^{-61}$). The red-dashed curve is for $d=1$, while the blue-dotted curve is for
$d=2$. The observational points correspond to the
$580$ SN Ia data points from \cite{Suzuki:2011hu}, and for
completeness and comparison  we depict the prediction of
$\Lambda$CDM cosmology with the black-solid curve. }} }
\label{Data1}
\end{figure}
\begin{figure}[ht]
\includegraphics[width=9.4cm,height=7cm]{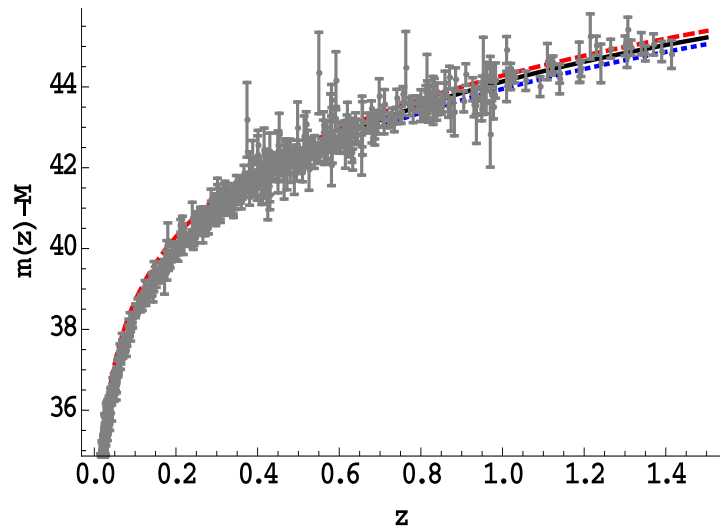}
\caption{ {\it{The theoretically
predicted apparent minus absolute magnitude as a function of the
redshift, for the logarithmic-corrected    fluid  (\ref{Eq4}) with the viscosity
 (\ref{Eq17b}) (i.e.  linear
time-dependent function $\xi_1(t)$), for $a_0=5$
 in units
where
$\kappa^2 = c = \hbar = 1$ (i.e where $H(z=0)\equiv H_0\approx 6
\times 10^{-61}$). The red-dashed curve is for
and $\tilde{c}=0.51\times 10^{-30}$,
 while the blue-dotted curve is for
 $\tilde{c}=0.55\times 10^{-30}$. The
observational points correspond to the
$580$ SN Ia data points from \cite{Suzuki:2011hu}, and for
completeness and comparison  we depict the prediction of
$\Lambda$CDM cosmology with the black-solid curve. }} }
\label{Data2}
\end{figure}

In the scenarios of the logarithmic-corrected    fluid  (\ref{Eq4}) of the present work,
the functions $H(t)$ and $a(t)$ are known, namely relations (\ref{Eq13}),(\ref{Eq14}) for
the case of viscosity (\ref{Eq9}) (i.e.  constant function $\xi_1(t)$), and relations
(\ref{Eq20}),(\ref{Eq21}) for
the case of viscosity (\ref{Eq17b}) (i.e.  linear
time-dependent function $\xi_1(t)$). Thus, $H(z)$ can be easily calculated since the
redshift is defined as $1+z=1/a$ in the case where we set the
present value of the scale factor to 1. Hence, for the case of the model with  viscosity
(\ref{Eq9}) we can find that
\begin{equation}
 H(z)=\sqrt{\frac{b}{d}}\sqrt{1+[a_0(1+z)]^{2d}},
\end{equation}
where according to the definitions below (\ref{Eq12}) the
corresponding fluid parameters are
\begin{eqnarray}
&&A=\frac{2b}{\kappa^2 \rho_*}\nonumber\\
&& \alpha=\frac{1}{3\kappa^2}\left(
\frac{2d}{3\rho_*}-\frac{2b}{\kappa^2\rho_*^2}-1
\right).
\end{eqnarray}
Similarly, for the case of the model with  viscosity
(\ref{Eq17b}) we can find that
\begin{equation}
 H(z)=\frac{3}{4}\tilde{c}^2\ln^2[a_0(1+z)],
\end{equation}
where as we mentioned $d_1= \tilde{c}/(3\kappa^2)$.

In Fig. \ref{Data1} we depict the theoretically predicted apparent minus absolute
magnitude  as a function of $z$, for the scenario of the logarithmic-corrected fluid
(\ref{Eq4}) with the viscosity  (\ref{Eq9}), as well as the prediction of $\Lambda$CDM
cosmology, on top of the $580$ SN Ia observational data points from \cite{Suzuki:2011hu}.
Similarly, in  Fig. \ref{Data2} we depict the same graphs, but for  the case of the
viscosity  (\ref{Eq17b}). As we can see, for both models the agreement with the SN Ia
data is excellent. We mention that this behavior arises from  both the logarithmic
correction as well as from the viscosity terms.
Interestingly enough, the scenario at hand is practically
indistinguishable from $\Lambda$CDM cosmology, although we have not considered an explicit
cosmological constant. This feature reveals the capabilities of the scenario.

\subsection{$H(z)$ probes}

In this subsection we will use the $H(z)$ Hubble function observations in order to impose
constraints on the free model parameters
\cite{Farooq:2016zwm,Capozziello:2017nbu,Capozziello:2018jya,Basilakos:2018arq}.
In particular, we will use the set  given in
\cite{Farooq:2016zwm}, which  contains $N=38$ entries in the  redshift interval
$0.07\leq z\leq 2.36$.
As it known,  the nominal chi-square function reads as \cite{Basilakos:2018arq}
\begin{equation}
\label{eq:xtetr}
    \chi^2_{H}(\phi^{\mu}) = {\mathbf V} {\mathbf C}^{-1}_{\text cov} {\mathbf V}^{T},
\end{equation}
with $\phi^{\mu}$   the statistical vector that contains
the free parameters,
${\mathbf C}^{-1}_{\text cov}$   the inverse of the covariance matrix
  and
$ {\mathbf V} = \{H_{D}(z_1) - H_{M}(z_1,\phi^{\mu}),...,H_{D}(z_{N}) -
H_{M}(z_N,\phi^{\mu})\}$.
Moreover, $z_{i}$ are the observed redshifts, while the letters $M$ and $D$
denote the data and models respectively. Hence,
the theoretical Hubble parameter is parametrized as
\begin{eqnarray}
\label{eq1}
H_{M}(z,\phi^{\mu}) = H_{0} E(z,\phi^{\mu+1}),
\end{eqnarray}
and therefore
\begin{equation}
\label{eq:Vec}
    {\mathbf V} = \{H_{D}(z_1) - H_{0}E(z_1,\phi^{\mu+1}),.., H_{D }(z_{N}) -
H_{0}E(z_N,\phi^{\mu+1})\},
\end{equation}
with $H_{0}$  the Hubble constant at present, $E(z)^2=H(z)^2/H_0^2$   the
dimensionless Hubble function,
and where the vector $\phi^{\mu+1}$
contains the model free parameters.

\begin{figure}[ht]
\includegraphics[width=10.cm,height=7.6cm]{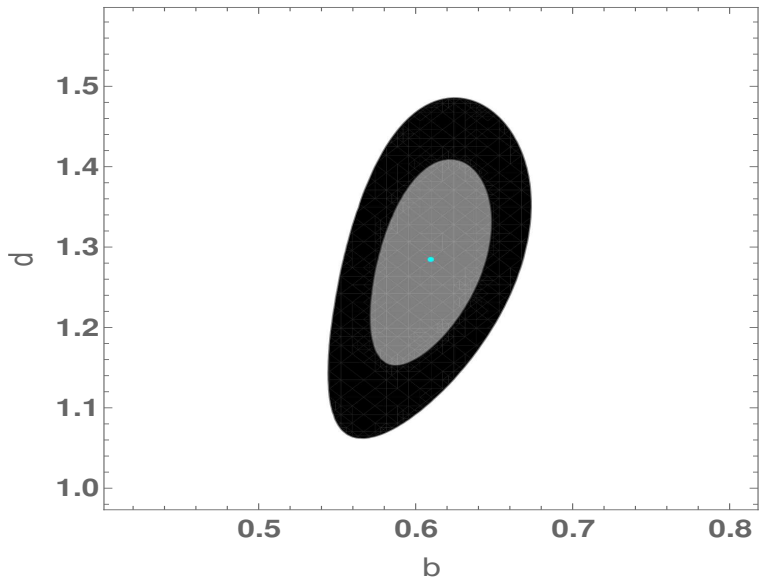}
\caption{ {\it{
Contour plots for the free parameters $b$ and $d$, for $a_0=1$, for   the model of    the
logarithmic-corrected  fluid  (\ref{Eq4}) with  viscosity
 (\ref{Eq9}).  The gray and black regions correspond respectively to 1$\sigma$ and
2$\sigma$ confidence level in the case of  $H(z)$ data sets of \cite{Farooq:2016zwm},
while the light blue dot marks the best-fit values.
 }} }
\label{logplot1}
\end{figure}

The usual way to proceed is to introduce the standard  $\chi^{2}$ estimator and
impose the exact value of $H_{0}$ ($H_{0} = 73.24 \pm 1.74$ Km/s/Mpc)
found by the SNIa team
(Riess et al. \cite{Riess:2016jrr}). However, and in order to bypass the
 $\sim 3\!-\!3.5\ \sigma$ tension with the
Planck Probe ($H_{0} = 67.8 \pm 0.9$ Km/s/Mpc \cite{Adam:2015rua}), we follow
\cite{Basilakos:2018arq} and we treat $H_{0}$ as a free parameter.

\begin{figure}[ht]
\includegraphics[width=10.cm,height=7.6cm]{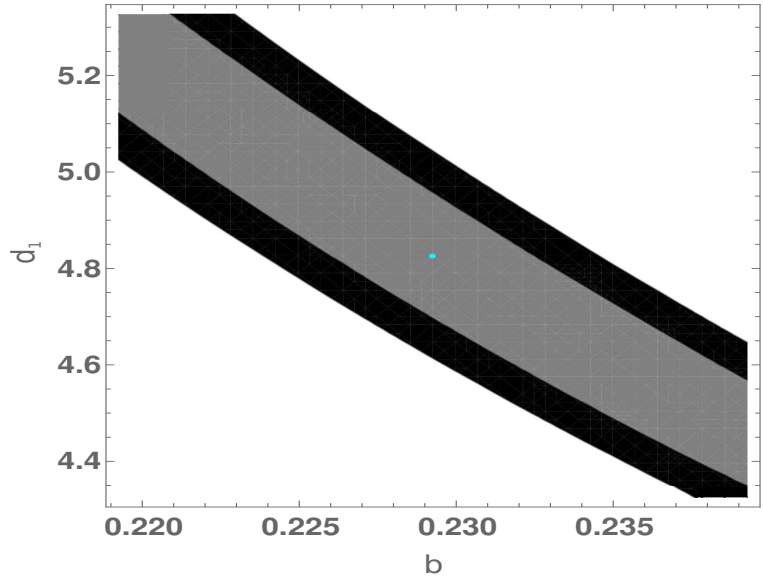}
\caption{ {\it{Contour plots for the free parameters $b_1$ and $d_1$, for free $a_0$,
for   the model of    the logarithmic-corrected  fluid  (\ref{Eq4}) with  viscosity
 (\ref{Eq17b}).  The gray and black regions correspond respectively to 1$\sigma$ and
2$\sigma$ confidence level in the case of  $H(z)$ data sets of \cite{Farooq:2016zwm},
while the light blue dot marks the best-fit values.
 }} }
\label{logplot2}
\end{figure}

In Fig. \ref{logplot1} we depict the  1$\sigma$ and
2$\sigma$  likelihood contours for the free parameters  $b$ and
$d$, as well as the corresponding best-fit values, for the model of the
logarithmic-corrected  fluid (\ref{Eq4}) with  viscosity  (\ref{Eq9}).
 Similarly, in  Fig. \ref{logplot2} we present the same graphs for the case of the
viscosity  (\ref{Eq17b}). As we observe, both models can be in agreement with
observations, which is an advantage of the scenario. Moreover, while in the first model
the free parameters are significantly constrained, in the second model the agreement with
the data can be obtained for a wide parameter range.

\section{Dynamical system analysis}
\label{DynamicalSystem}

In this section we apply the powerful method of dynamical system analysis in order
to investigate the full system of cosmological equations. This method allows   to
extract information about the global behavior of a cosmological scenario, bypassing the
complexity of the involved equations
\cite{Leon2011,Bahamonde:2017ize,Copeland:1997et,Leon:2009rc,Leon:2010pu,
Kofinas:2014aka,Odintsov:2017tbc, Oikonomou:2017ppp,Kleidis:2018cdx}. In order to be
fully general, in the following we will assume an integration between the dark energy
and the matter fluids, which is widely used since it can provide   an alleviation to the
coincidence
problem while it cannot be theoretically excluded
\cite{Gondolo:2002fh,Farrar:2003uw,Cai:2004dk,Wang:2006qw,
Bertolami:2007zm,Chen:2008ft,
He:2008tn,Valiviita:2008iv,Jackson:2009mz,Jamil:2009eb,He:2010im,
Boehmer:2008av,Bamba:2012cp,Bolotin:2013jpa,
Costa:2013sva,Nunes:2016dlj,Odintsov:2018awm,Odintsov:2018uaw}.

Our aim is to investigate the phase structure of a coupled dark
energy system, with the dark energy sector having an
EoS which has a logarithmic dependence and  bulk viscosity. In this case the two
Friedmann equations write as
\begin{equation}\label{flateinstein}
H^2=\frac{\kappa^2}{3}\rho_{tot}\, ,
\end{equation}
\begin{equation}\label{derivativeofh}
\dot{H}=-\frac{\kappa^2}{3}\left(
\rho_{tot}+p_{tot}\right)\, ,
\end{equation}
where $\rho_{tot}=\rho_d+\rho_m$ and $p_{tot}=p_d+p_m$,
with $\rho_d$ and $\rho_m$ the energy densities and $p_d$ and $p_m$ the pressures of the
two fluids. Concerning the matter fluid we assume it to be pressureless, namely $p_m=0$,
while for the dark energy
fluid we consider the logarithmic EoS with bulk
viscosity
\begin{equation}\label{darkenergyeos}
p_d=A\kappa^2\rho_d \ln (\kappa^2 \rho_d)-A\kappa^2\rho_d \ln
(3H^2)\, ,
\end{equation}
where $A$ is the model parameter. In the interacting scenario at hand the conservation
equations for the two fluids are written as
\begin{align}\label{continutiyequations}
& \dot{\rho}_m+3H\rho_m=-Q\,
\\ \notag & \dot{\rho}_d+3H(\rho_d+p_d)=Q\, ,
\end{align}
with $Q$ being the interaction between the dark fluids. The algebraic sign of the
interaction term $Q$ determines which fluid loses energy, thus if
$Q>0$ it implies
that dark matter sector loses energy towards the dark
energy one. We shall use a phenomenologically motivated
interaction term of the form
\cite{CalderaCabral:2008bx,Pavon:2005yx,Quartin:2008px,Sadjadi:2006qp,Zimdahl:2005bk},
\begin{equation}\label{qtermform}
Q=3H(c_1\rho_m+c_2\rho_d)\, ,
\end{equation}
with $c_1$, $c_2$  real constants.

Having Eqs. (\ref{flateinstein}), (\ref{derivativeofh}) and
(\ref{continutiyequations}) at hand we construct an autonomous dynamical system  by
choosing the dimensionless variables
\begin{equation}\label{variablesofdynamicalsystem}
x_1=\frac{\kappa^2\rho_d}{3H^2},\,\,\,x_2=\frac{\kappa^2\rho_m}{3H^2},\,\,\,z=\kappa^2H^2
\,.
\end{equation}
The variables of the dynamical system $x_1$ and $x_2$ satisfy the
Friedmann constraint,
\begin{equation}\label{friedmannconstraint}
x_1+x_2=1\, ,
\end{equation}
which is essentially the Friedmann equation (\ref{flateinstein}).
Additionally, the total equation of state parameter $w_{eff}$, defined as
\begin{equation}\label{totelesosparm}
w_{eff}=\frac{p_d}{\rho_{tot}}\, ,
\end{equation}
can easily be expressed in terms of the dynamical system
parameters $x_1$, $x_2$ and $z$ (\ref{variablesofdynamicalsystem})
in the following way:
\begin{equation}\label{equationofstatetotal}
w_{eff}=\tilde{A}x_1\ln x_1\, ,
\end{equation}
with
\begin{equation}\label{tildea}
\tilde{A}=A\kappa^2\, .
\end{equation}
Moreover, the interaction term  $Q$ appearing in (\ref{qtermform}), expressed in
terms of the variables $x_1$, $x_2$ and $z$
(\ref{variablesofdynamicalsystem}) is written as
\begin{equation}\label{additionalterms}
\frac{\kappa^2Q}{3H^3}=3c_1x_2+3c_2x_1\, .
\end{equation}

By combining Eqs. (\ref{flateinstein}), (\ref{derivativeofh}),
(\ref{continutiyequations}), (\ref{additionalterms}) and
(\ref{variablesofdynamicalsystem}), and also by using the
$e$-foldings number $N$ as a dynamical variable instead of the
cosmic time $t$,  we obtain the following
dynamical system:
\begin{align}
&\notag \frac{\mathrm{d}x_1}{\mathrm{d}N}=3 \tilde{A} (x_1-1) x_1 \ln
(x_1)-(c_1 x_2+c_2 x_1)+3 x_1^2+3 x_1 (x_2-1)\, ,
\\ \notag &
\frac{\mathrm{d}x_2}{\mathrm{d}N}=3 \tilde{A} x_1 x_2 \log
(x_1)+c_1 x_2+c_2 x_1+3 (x_1-1) x_2+3 x_2^2\, ,
\\  &
\frac{\mathrm{d}z}{\mathrm{d}N}=-3 \tilde{A} x_1 z \log (x_1)-3
x_1 z-3 x_2 z\, .\label{dynamicalsystemmultifluid}
\end{align}
The fixed points  $\phi_*^1=(x_1,x_2,z)$ of the above dynamical system, which are
obtained by setting the left-hand-sides of the equations to zero, cannot easily be
found analytically, and therefore we will investigate the phase space
behavior by solving numerically the dynamical system and by using
various initial conditions. Additionally, the stability of the fixed points can
be investigated by calculating explicitly the Jacobian of the
dynamical system at the fixed point $\phi_*^1$. The Jacobian
matrix, which we denote as $\mathcal{J}$, corresponds to the
linearized dynamical system near the resulting fixed point, namely
\begin{equation}\label{jaconiab}
\mathcal{J}=\sum_i\sum_j\Big{[}\frac{\mathrm{\partial
f_i}}{\partial x_j}\Big{]}\, .
\end{equation}
The Jacobian matrix $\mathcal{J}$ must be evaluated exactly at the
fixed points, and the corresponding eigenvalues   indicate
whether the particular fixed point is stable or not, whenever the fixed point
is hyperbolic (the
eigenvalues of the Jacobian at the fixed point have real parts
which are non-zero).
 In the case at hand, the Jacobian matrix of
the dynamical system (\ref{dynamicalsystemmultifluid}) is
\begin{equation}\label{jacobianofdynsystem}
\mathcal{J}=\left(
\begin{array}{ccc}
 -24 x_1+3 x_2+(30-60 x_1) \log (x_1)+26 & 3 x_1-1 & 0 \\
 -30 \log (x_1) x_2-27 x_2+1 & -30 \log (x_1) x_1+3 x_1+6 x_2-2 & 0 \\
 3 z (10 \log (x_1)+9) & -3 z & 30 x_1 \log (x_1)-3 (x_1+x_2) \\
\end{array}
\right)\, .
\end{equation}

A thorough investigation of the parameter space reveals that
the sign of the parameter $\tilde{A}$, and also the sign of the
parameters $c_1$ and $c_2$, crucially affect the behavior of the
dynamical system. Actually, for a wide range of values, the
dynamical system is strongly unstable, however there are some
regions of the parameter space for which stability occurs. These
stability regions are the most interesting ones, from the physical point
of view.
\begin{table*}[h]
\begin{tabular}{@{}crrrrrrrrrrrrrrrrrrr@{}}
\tableline
 Case I: $c_1>0$, $c_2>0$, and $\tilde{A}>0$:  Strong Instability of Phase Space,
Variables Blow-up.
\\\tableline
Case II: $c_1<0$, $c_2<0$, and $\tilde{A}>0$:  Strong Instability
of Phase Space, Variables Blow-up.
\\\tableline
Case III: $c_1>0$, $c_2=0$, and $\tilde{A}>0$:  Strong Instability
of Phase Space, Variables Blow-up.
\\\tableline
Case IV: $c_1=0$, $c_2>0$, and $\tilde{A}>0$:  Strong Instability
of Phase Space, Variables Blow-up.
\\\tableline
 Case V: $c_1=0$, $c_2<0$, and $\tilde{A}>0$:
Strong Instability of Phase Space, Variables Blow-up.
\\\tableline
 Case VI: $c_1<0$, $c_2=0$, and $\tilde{A}>0$:
Strong Instability of Phase Space, Variables Blow-up.
\\\tableline
Case VII: $c_1>0$, $c_2=0$, and $\tilde{A}>0$: Strong Instability
of Phase Space, Variables Blow-up.
\\\tableline
Case VIII: $c_1<0$, $c_2<0$, and $\tilde{A}<0$: Strong Instability
of Phase Space, Variables Blow-up.
\\\tableline
Case IX: $c_1>0$, $c_2>0$, and $\tilde{A}<0$: Stable Fixed Point.
\\\tableline
Case X: $c_1>0$, $c_2=0$, and $\tilde{A}<0$: Stable Fixed Point.
\\\tableline
Case XI: $c_1=0$, $c_2>0$, and $\tilde{A}<0$: Stable Fixed Point.
\\\tableline
\end{tabular}
\small \caption{\label{numericalanalysisofsystem}
Stability regions of the
phase space of the dynamical system corresponding to the
two-fluids Universe (\ref{dynamicalsystemmultifluid}).}
\end{table*}
\begin{figure}[h]
\centering
\includegraphics[width=18pc]{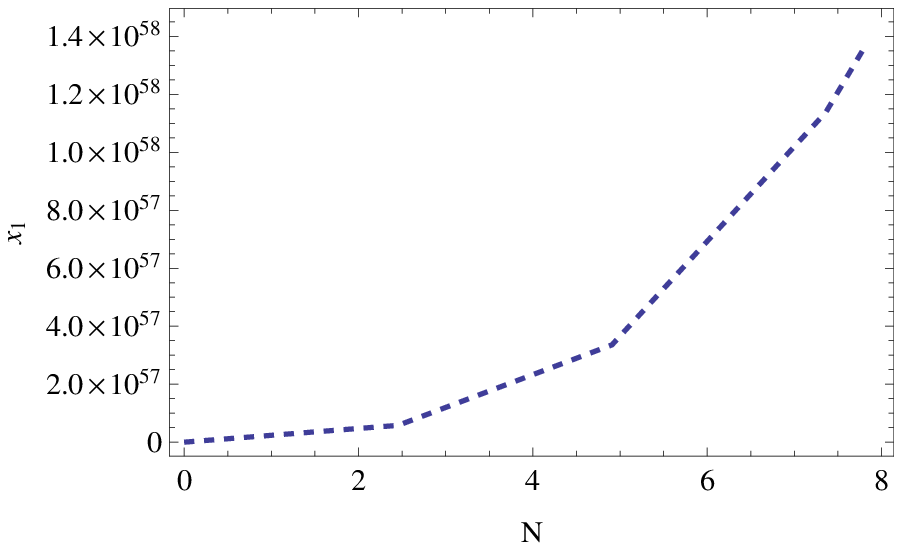}
\includegraphics[width=18pc]{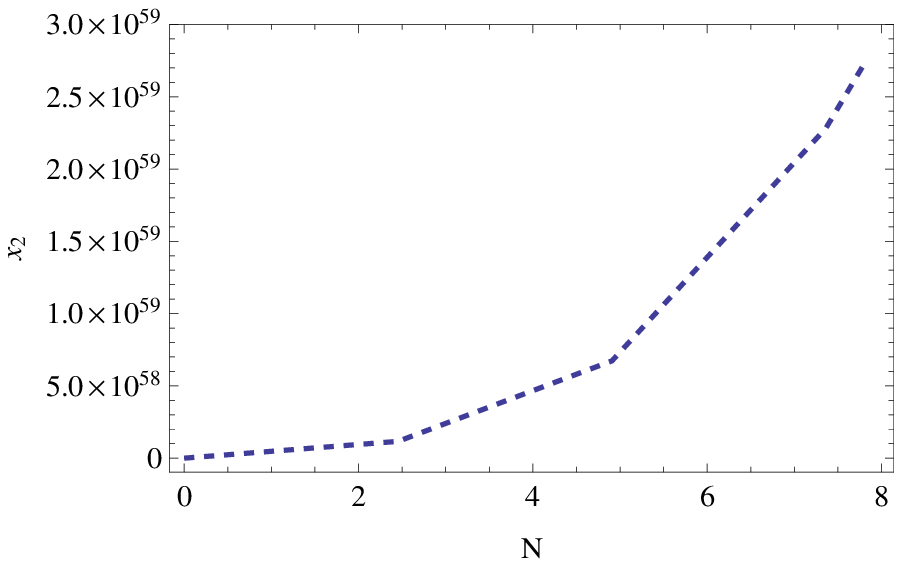}
\includegraphics[width=18pc]{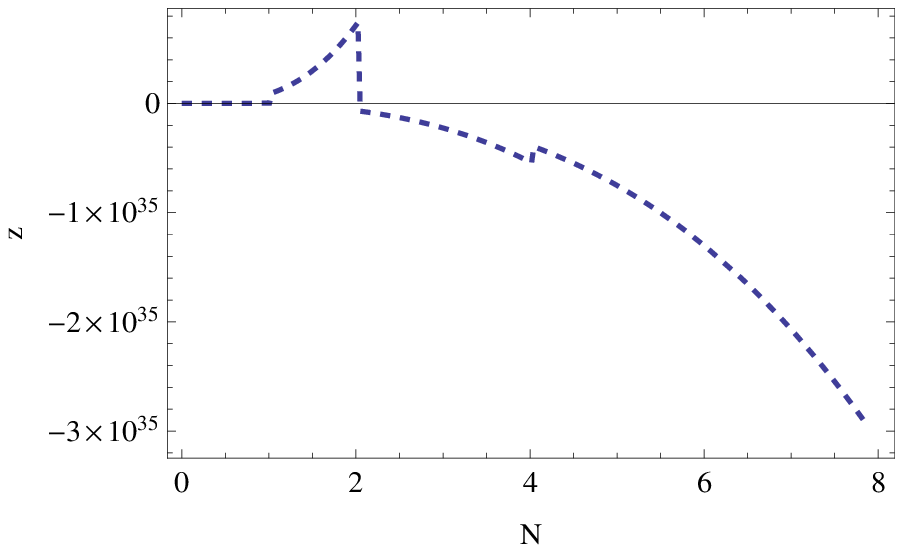}
\caption{{\it{The evolution   of the variables $x_1(N)$
(left), $x_2(N)$ (right) and $z(N)$ (bottom) as functions of the
$e$-foldings number $N$, for $c_1=c_2=1$ and $\tilde{A}=10$.}}}
\label{plot2}
\end{figure}
\begin{figure}[h]
\centering
\includegraphics[width=18pc]{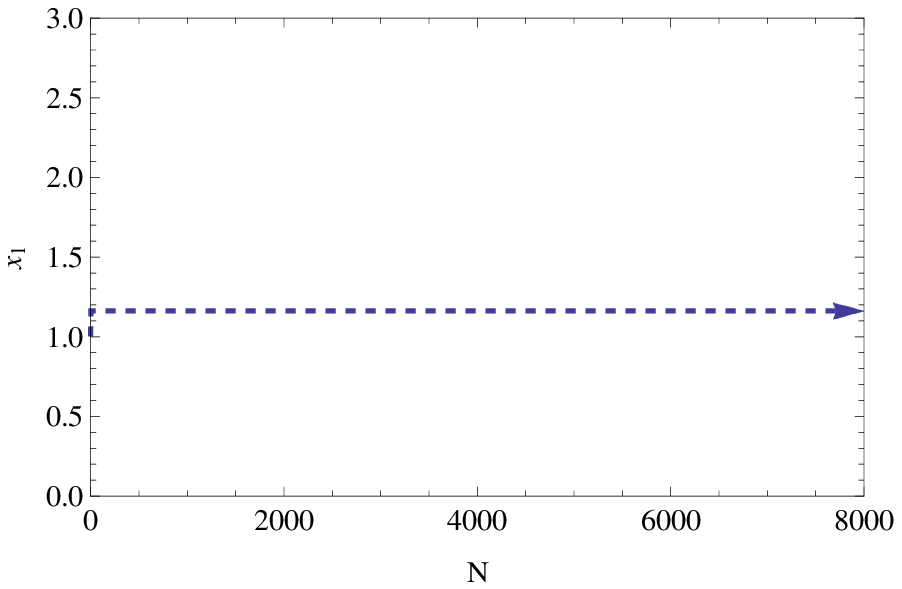}
\includegraphics[width=18pc]{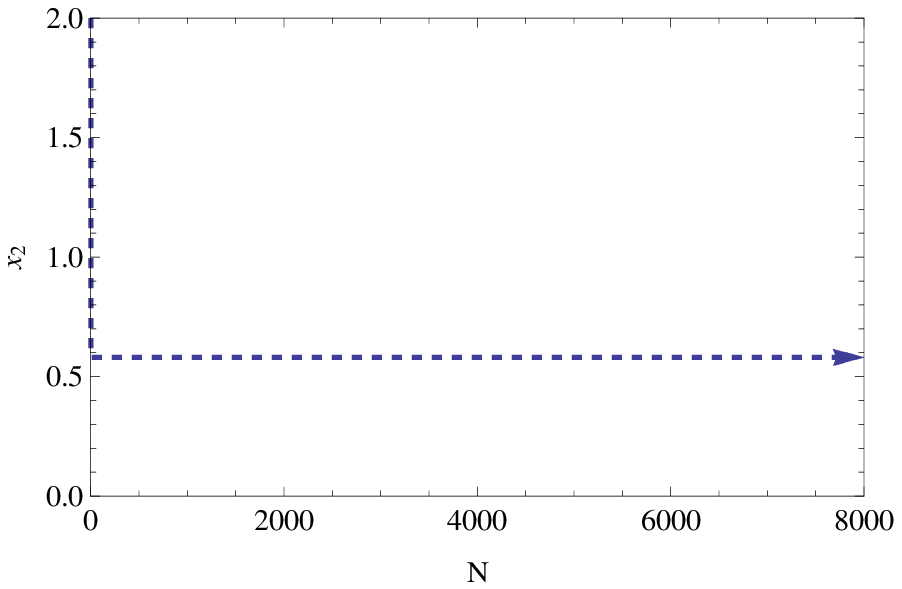}
\includegraphics[width=18pc]{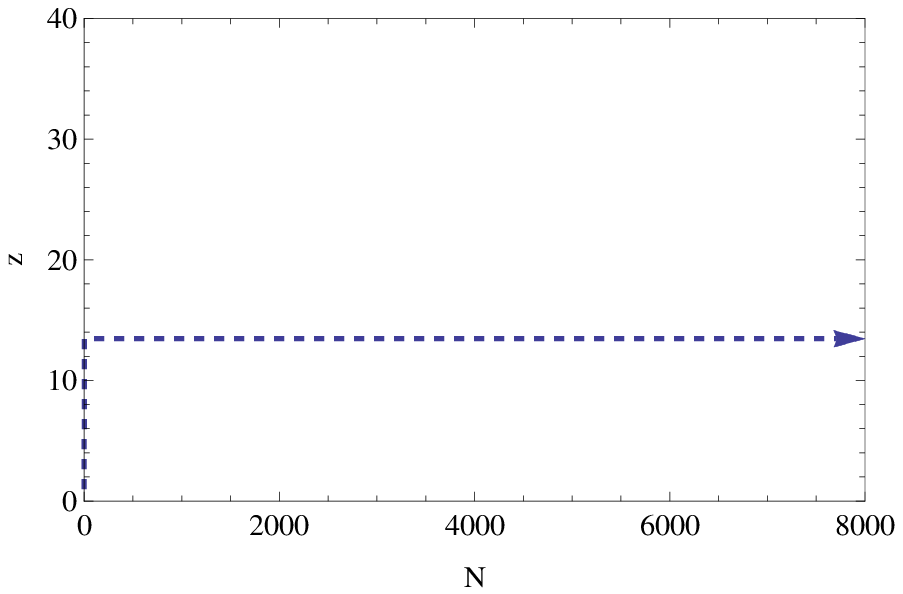}
\caption{{\it{The evolution of the variables $x_1(N)$
(left), $x_2(N)$ (right) and $z(N)$ (bottom) as functions of the
$e$-foldings number $N$, for $c_1=c_2=1$ and $\tilde{A}=-10$.}}}
\label{plot3}
\end{figure}

In Table
\ref{numericalanalysisofsystem} we display the results of our numerical
analysis with regard to the stability regions. From the phenomenological point of view,
a stable fixed point is physically important since it can attract
the Universe at late times. Hence,  as we observe from Table
\ref{numericalanalysisofsystem}, the most important
are the cases IX, X and XI. Before analyzing them in details, let us first have a
clear picture of the instability of
the phase space for the rest of the cases. In Fig. \ref{plot2} we
plot the functional dependence of the variables $x_1(N)$ (left),
$x_2(N)$ (right) and $z(N)$ (bottom) in terms of the
$e$-foldings number $N$, for $c_1=c_2=1$ and $\tilde{A}=10$. The
late-time behavior is achieved for large values of the
$e$-foldings number $N$. As it can be seen, even from the first
$e$-foldings no equilibrium (fixed point) is reached, and all the
variables blow-up at finite-time. This behavior occurs always for
all positive $\tilde{A}$ values, regardless of the values of the
parameters $c_i$, $i=1,2$.

\begin{figure}[h]
\centering
\includegraphics[width=20pc]{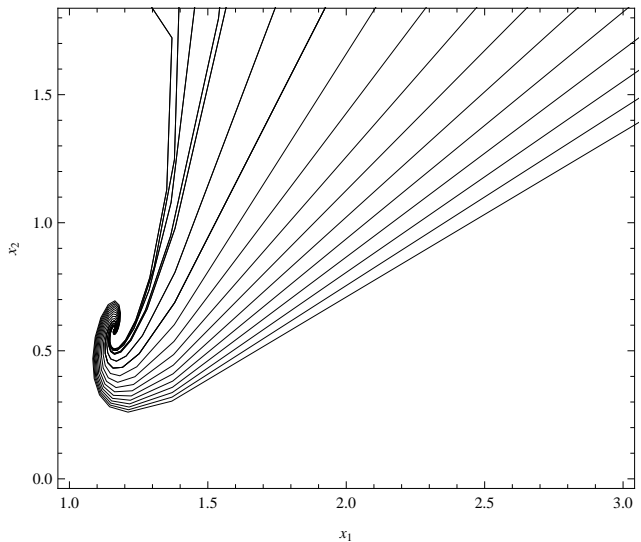}
\caption{{\it{The trajectories in the projected phase space $x_1-x_2$ for
various initial conditions, and for $c_1=c_2=1$ and $A=-10$. The resulting stable fixed
point corresponds to case IX of Table \ref{numericalanalysisofsystem}.}}}
\label{plot5}
\end{figure}

Having discussed the instability regions, let us now proceed to the stable regimes,
focusing on the IX case in Table \ref{numericalanalysisofsystem}, in
which   $c_1>0$, $c_2>0$, and $\tilde{A}<0$. In Fig. \ref{plot3} we plot
the functional dependence of the variables $x_1(N)$ (left),
$x_2(N)$ (right) and $z(N)$ (bottom) in terms of the
$e$-foldings number $N$, for $c_1=c_2=1$ and $A=-10$.  As it can be seen, for large
values of $N$ a stable
fixed point is reached (its behavior does not
change with $N$), regardless
the initial conditions used for the variables $x_1(N)$, $x_2(N)$
and $z(N)$.
As can be verified numerically, the stable fixed point of this specific example
is
\begin{equation}\label{fixedpoint1}
\phi_*^1=(x_1,x_2,z)=(1.16183,0.580917,13.794)\, ,
\end{equation}
 and thus the eigenvalues of the  Jacobian matrix $\mathcal{J}$ of
(\ref{jacobianofdynsystem})    at this fixed point are
\begin{equation}\label{eigenvalues}
(-3.17751+5.87099 i,-3.17751-5.87099 i,-1.77636\times 10^{-15})\,.
\end{equation}
Hence, due to the fact that the eigenvalues (\ref{eigenvalues})
have negative real parts,  the fixed point
(\ref{fixedpoint1}) is a stable hyperbolic fixed point. It
is necessary to  examine the stability of the fixed point for various
initial conditions, and thus in Fig. \ref{plot5} we present the projected phase-space
plots for various initial conditions, in the case where $c_1=c_2=1$
and $A=-10$. As it can be seen, all the trajectories in the phase
space $x_1-x_2$ tend asymptotically to the fixed point. Actually,
the fixed point is reached quite fast, as it can be easily
checked.

\begin{figure}[h]
\centering
\includegraphics[width=18pc]{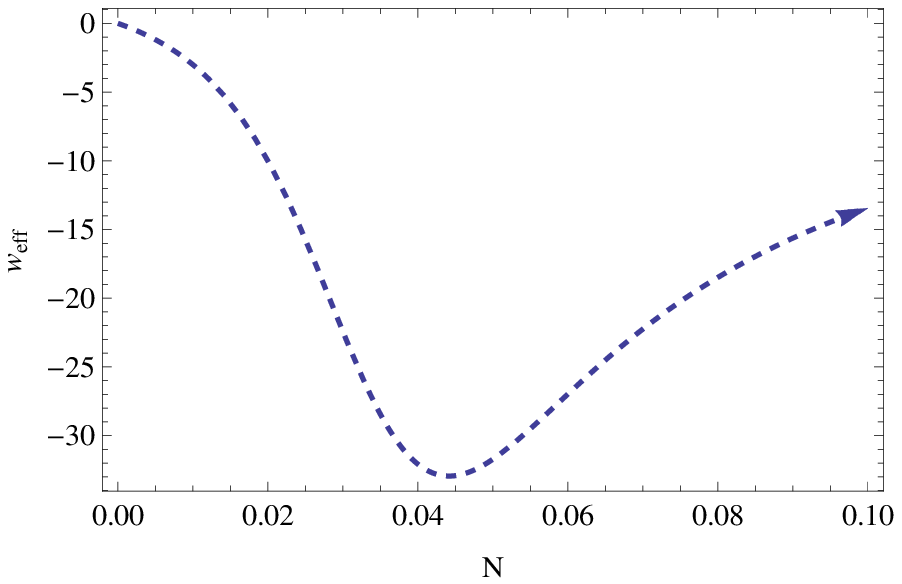}
\includegraphics[width=18pc]{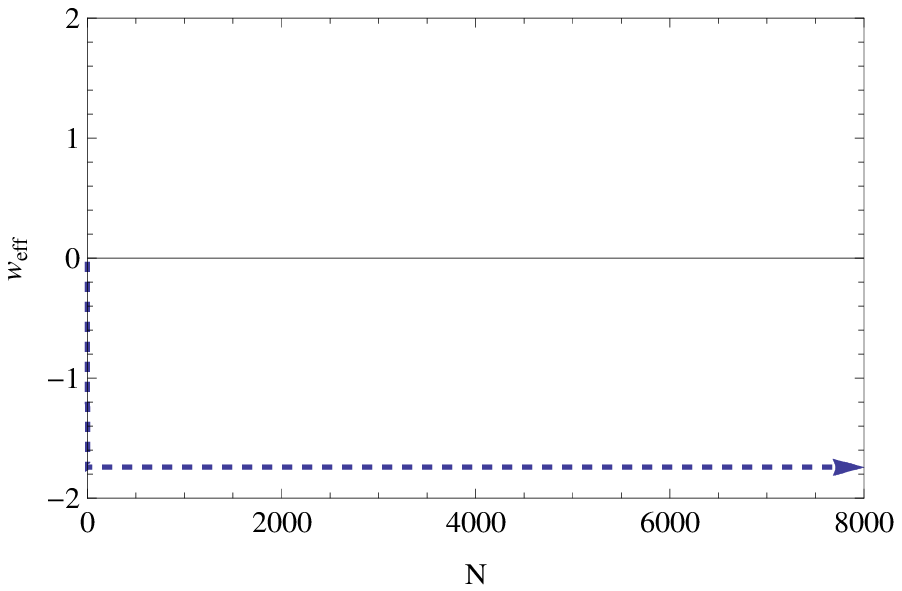}
\caption{{\it{The evolution of the total equation of
state parameter $w_{eff}$ of
(\ref{equationofstatetotal}), as a function of $N$, for $c_1=c_2=1$
and $\tilde{A}=-10$.}}} \label{plot4}
\end{figure}

In order to present the physical behavior of the above solution in a more transparent
way, in Fig. \ref{plot4} we depict the evolution of the total equation-of-state
parameter
 $w_{eff}$ given in (\ref{equationofstatetotal}), corresponding to the above numerical
example. As it can be seen, the resulting EoS indicates an accelerating Universe at late
times. A noticeable behavior occurs for large negative values of
$\tilde{A}$, and for small and positive values of the parameters
$c_1$ and $c_2$, in which case  the EoS parameter
tends asymptotically to a de Sitter value $w_{eff}\sim -1$, and thus the dynamical dark
energy experiences
a stabilization towards the cosmological constant value due to the logarithmic
correction and interaction term.

In summary, the addition of the logarithmic corrections in the dark energy equation of
state, leads either to instabilities in the phase space, or to a stable phantom fixed
point at late times, and therefore to late-time acceleration. The behavior of the
dynamical system for the cases X and XI is similar to the case IX, and thus we omit it for
brevity. It is important to note that this model is just a simple model containing a
logarithmic dependence, and in principle a more viable model can be achieved by
appropriately modifying the EoS. In particular, more realistic equations of state for the
dark energy fluid can be chosen, for example including polynomial terms along with the
logarithmic correction.

\section{Conclusions}
\label{Conclusions}

In this work we have studied  dark energy scenarios which involve equation of states with
logarithmic terms of the energy density, allowing additionally for a bulk viscosity. The
logarithmic-corrected power-law fluid possesses properties analogous to properties of
isotropic deformation of crystalline solids, in which the pressure can be
negative. Without loss of generality, we considered two classes of the model, arisen from
two different  viscosity considerations.
In both  scenarios we were able to obtain the accelerating expansion of
the late Universe, since  we acquired the transition from a decelerating phase
to an accelerating one, without the explicit consideration of a cosmological constant.

In order to obtain a better picture of the phenomenology of the construction, we
performed a confrontation with observations, and in particular with the Supernovae type
Ia (SN Ia) and $H(z)$ Hubble function  data, constructing the model parameter contour
plots. As we showed, for both models the agreement with the is excellent, as a result of
both the logarithmic correction as well as of the
viscosity terms.

Additionally, we applied the powerful method of dynamical system analysis in order to
extract information about the global behavior of the cosmological scenario at hand,
bypassing the complexity of the involved equations, where for generality we also
considered an interaction between the dark energy and   matter fluids. By
appropriately choosing the dimensionless variables we constructed an autonomous
dynamical system, and we investigated its dynamical evolution  for various
parameter
values and for several initial conditions. As we showed, there
exist asymptotic accelerating attractors, which in most cases
correspond to a phantom late-time evolution. However, in some
cases it is possible to obtain a nearly de Sitter late-time
evolution by appropriately choosing the model parameters, which is a significant
capability of the scenario since the dark energy, although dynamical, stabilizes at the
cosmological constant value.

In summary, fluids with logarithmic equation of state may lead to interesting
cosmological behavior, and thus they can be a good candidate for the description of the
dark energy sector. Definitely, an important and necessary investigation is to analyze in
detail the cosmological perturbations, and confront them with the observed large scale
structure behavior. However, this study lies outside the scope of the present work, and
it is left for a future project.

\section*{Acknowledgments}
The authors wish to thank S. Pan and W. Yang for useful
discussions. This work is supported by MINECO (Spain),
FIS2016-76363-P (S.D.O), by project 2017 SGR247 (AGAUR, Catalonia)
(S.D. Odintsov). This article is based upon work from COST Action
``Cosmology and Astrophysics Network for Theoretical Advances and
Training Actions'', supported by COST (European Cooperation in
Science and Technology). This research was also supported in part
by Russian Ministry of Education and Science, project No.
3.1386.2017 (S.D. Odintsov).

\end{document}